\documentclass[iop,apjl]{emulateapj}
\usepackage{epsfig}
\usepackage{amsmath, amsthm, amssymb}
\usepackage[plainpages=false, colorlinks=true, anchorcolor=blue, linkcolor=blue, citecolor=blue, bookmarks=false, urlcolor=blue]
{hyperref}
\usepackage{color}
\usepackage{microtype}
\usepackage{float}
\usepackage{verbatim}
\usepackage{wrapfig}
\usepackage{graphicx}
\usepackage[caption = false]{subfig}
\usepackage{multirow}
\usepackage{hhline}
\usepackage{booktabs}
\usepackage{array}

\newcommand{\Msun}{$M_\sun$}

\newcommand\Tstrut{\rule{0pt}{3.0ex}}

\newcommand{\shortauth}{S. Swihart}
\newcommand{\slugcom}{Accepted to ApJ - March 20, 2019}
\slugcomment{\slugcom}

\lefthead{\sc \footnotesize \slugcom \hfill \shortauth}
\righthead{\sc \footnotesize \slugcom \hfill \shortauth}

\begin{document}

\title{PSR J1306--40: An X-ray Luminous Redback with an Evolved Companion}

\author{Samuel J. Swihart\altaffilmark{1},
Jay Strader\altaffilmark{1},
Laura Chomiuk\altaffilmark{1},
Laura Shishkovsky\altaffilmark{1}
}

\affil{ 
  \altaffilmark{1}{Center for Data Intensive and Time Domain Astronomy, Department of Physics and Astronomy,\\Michigan State University, East Lansing, MI 48824, USA}\\
}

\begin{abstract}
PSR J1306--40 is a millisecond pulsar binary with a non-degenerate companion in an unusually long $\sim$1.097 day orbit. We present new optical photometry and spectroscopy of this system, and model these data to constrain fundamental properties of the binary such as the component masses and distance. The optical data imply a \emph{minimum} neutron star mass of $1.75\pm0.09\,M_{\odot}$~(1-sigma) and a high, nearly edge-on inclination. The light curves suggest a large hot spot on the companion, suggestive of a portion of the pulsar wind being channeled to the stellar surface by the magnetic field of the secondary, mediated via an intrabinary shock. The H$\alpha$ line profiles switch rapidly from emission to absorption near companion inferior conjunction, consistent with an eclipse of the compact emission region at these phases. At our optically-inferred distance of $4.7\pm0.5$ kpc, the X-ray luminosity is $\sim$10$^{33}$ erg s$^{\textrm{-1}}$, brighter than nearly all known redbacks in the pulsar state. The long period, subgiant-like secondary, and luminous X-ray emission suggest this system may be part of the expanding class of 
millisecond pulsar binaries that are progenitors to typical field pulsar--white dwarf binaries.
\end{abstract}

\section{Introduction}
Millisecond pulsars (MSPs) form in binary systems, where the central neutron star accretes matter and angular momentum from a non-degenerate companion, spinning it up to very rapid spin periods. Although this mass transfer process is expected to last hundreds of Myrs or more \citep{Tauris99}, most MSPs in the Galactic field have degenerate white dwarf companions in wide orbits ($P_{\textrm{orb}}\gtrsim$ 5 d), which represent the end stage of the recycling process \citep{Tauris06}.

Recent multiwavelength (optical, radio, and X-ray) follow-up observations of unidentified \emph{Fermi} Large Area Telescope (LAT) $\gamma$-ray sources have revealed a substantial number of close neutron star binary systems that host hydrogen-rich secondaries rather than He white dwarfs. Unlike most field MSPs (as described above), these appear to be systems in which the standard recycling process is not yet complete \citep[e.g.,][]{Benvenuto14}, providing valuable insights into the spin-up process. These binaries are classified by the mass of the secondary: ``redbacks"  have non-degenerate, main sequence-like companions ($M_c \gtrsim$ 0.1 \Msun), compared to the less massive ($M_c \lesssim$ 0.05 \Msun), highly ablated, semi-degenerate ``black widows'' \citep{Roberts11b}. Both systems show radio eclipses due to ionized material from the secondary; these eclipses are typically more extensive for redbacks \citep[e.g.,][]{DAmico01, Camilo15, Cromartie16}.

A few redbacks have been observed to transition back and forth between an accretion-powered disk state and a rotationally-powered pulsar state. These ``transitional millisecond pulsars'' proved the suspected evolutionary link between some recycled MSPs and neutron star low-mass X-ray binaries \citep[e.g.,][]{Archibald09,Papitto13,Bassa14}. However, like the bulk of the redback population, the short orbital periods in these systems ($\lesssim$ 0.5 d) mean they are unlikely to end their lives as the wide-orbit MSP--He white dwarf binaries that dominate the observed population of MSPs in the field \citep{Chen13}. Instead, the progenitors to these canonical MSP--white dwarf systems are neutron stars with evolved red giant companions whose orbital periods are $>$ 1 day \citep{Tauris99}.

In this context, the MSP binary 1FGL J1417.7--4407 (PSR J1417--4402) was the first MSP binary discovered that is a likely progenitor to the \emph{typical} MSP-white dwarf binaries \citep{Strader15,Camilo16}. Due to its long period, red giant companion, and inferred evolutionary track, a new ``huntsman'' subclass was coined to distinguish unique systems like these from the more common redbacks \citep{Strader15, Swihart18}.

Optical spectra of 1FGL J1417.7--4407 show a strong, persistent double-peaked H$\alpha$ emission profile that is unusual among redbacks in their pulsar states and reminiscent of a classic accretion disk. However, due to the small separation between the peak components, and the stationary profiles as a function of orbital phase in the rest frame of the secondary, this H$\alpha$ phenomenology is likely not due to a disk. Instead, the emission likely comes from a combination of material swept off the companion and a strong intrabinary shock \citep{Camilo16,Swihart18}. Another piece of evidence for an unusually luminous shock is the X-ray luminosity in this system ($\sim$ 10$^{33}$ erg s$^{\textrm{-1}}$), which is higher than other redbacks in the pulsar state. \citet{Swihart18} discuss the possibility that the shock luminosity in 1FGL J1417.7--4407 is enhanced over typical redbacks by a strong, magnetically driven wind from the tidally-locked red giant secondary. The candidate MSP binary 2FGL J0846.0+2820 shows many similarities to 1FGL J1417.7--4407, making it a possible second member of the huntsman subclass, though no radio pulsar has yet been confirmed in this system \citep{Swihart17}. Pending future evolution studies, it might be reasonable to include other long period redback-like systems with evolved companions in this class (e.g., PSR NGC 6397A).

The subject of this paper, PSR J1306--40, was discovered in the SUPERB survey \citep{Keane17}, where it was identified as a candidate redback binary. No pulsar timing solution was presented owing to the frequent eclipses. \citet{Linares18} found the X-ray spectrum of this source is a hard powerlaw typical of redbacks, and that the X-ray and optical light curves are modulated on the same orbital period, confirming the source as a likely redback. The orbital period is one of the longest known among compact MSP binaries ($P_{\textrm{orb}}\sim1.097$ d).

Here we present the first optical spectroscopy and multi-band optical photometry of PSR J1306--40, and show that it has properties more similar to huntsman-type MSP binaries than classic redback systems. We detail our new optical spectroscopic and photometric observations of the system in \S\ref{sec:optobs}. We model the optical light curves in \S\ref{sec:LCfitting}, and examine the phase-resolved H$\alpha$ profiles in \S\ref{sec:Halpha}. Finally, we make concluding remarks and discuss this system in the context of known redbacks and its connection to MSP--white dwarf progenitors in \S\ref{sec:discussion}.

\section{Optical Observations}
\label{sec:optobs}

\subsection{SOAR Spectroscopy}
\label{sec:spectroscopy}
We obtained spectra of the companion to PSR J1306--40 using SOAR/Goodman \citep{Clemens04} from 2017 Jul 11 to 2018 Jun 9. All data used a 1200 l mm$^{-1}$ grating and a 0.95\arcsec\ slit, giving a resolution of about 1.7 \AA. Exposure times ranged from 15 to 30 min per spectrum, depending on weather conditions. The spectra  covered a wavelength range of $\sim$ 5485--6740 \AA. Barycentric radial velocities were determined through cross-correlation with bright standards taken with the same setup, primarily in the wavelength range 6050--6250 \AA. The resulting 43 velocities are listed in Table~\ref{tab:RVdata}. Observation epochs have been corrected to Barycentric Julian Date (BJD) on the Barycentric Dynamical Time system \citep{Eastman10}.

\begin{figure}[t]
\includegraphics[width=3.5in,trim={0 5mm 0 0},clip]{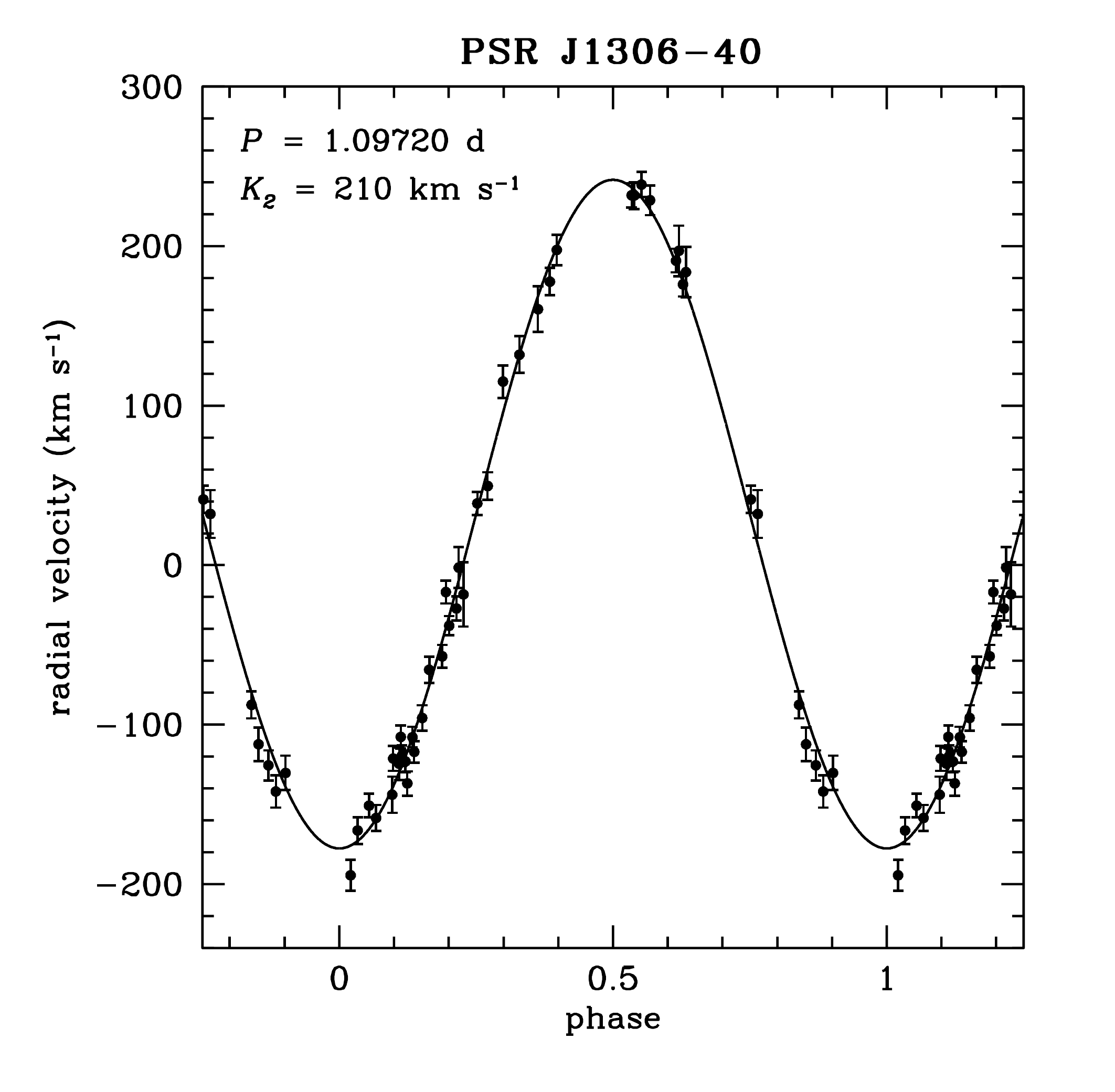}
\caption{Circular Keplerian fit to the SOAR/Goodman barycentric radial velocities of PSR J1306--40 listed in Table~\ref{tab:RVdata}.}
\label{fig:j1306_rv}
\vspace{4mm}
\end{figure}

Even though the SUPERB survey has detected a pulsar in this system \citep{Keane17}, it has not yet been timed, so the ephemerides must be determined from our data. Using the custom Markov Chain Monte Carlo sampler \emph{TheJoker} \citep{Price17}, we fit a circular Keplerian model to the data in Table~\ref{tab:RVdata} in order to determine the orbital period $P$, BJD time of the ascending node $T_0$, systemic velocity $\gamma$, and the semi-amplitude $K_{2}$. Here, and throughout the paper, uncertainties are given at 1-sigma. We find $P = 1.097195(161)$ d, $T_0 = 2457780.8373(19)$ d, $\gamma = 32.0\pm1.8$ km s$^{-1}$, $K_{2} = 210 \pm 2$ km s$^{-1}$. (We note that the spectroscopic period we find here is fully consistent with the photometric period found in the discovery paper \citet{Linares18}). A fit using these median values is shown in Figure~\ref{fig:j1306_rv}. For the remainder of this work, we assume the period derived from our spectroscopy.

This fit has an rms scatter of 11 km s$^{-1}$ (compared to a median uncertainty among the velocities of about 8.4 km s$^{-1}$) and a $\chi^2$/d.o.f of 74/39, perhaps suggesting the fit could be improved. However, we see no clear trends in the residuals. A fit with the eccentricity left free does not find a value significantly different from 0, so there is no evidence that the orbit is non-circular. It may just be that the velocity uncertainties are slightly underestimated. It would be useful to obtain additional radial velocities in the range $\phi = 0.7-1.0$; it is challenging to get complete phase coverage for this system since its period is close to 1 d. Future pulsar timing should improve the ephemerides by orders of magnitude.

Our SOAR spectra also show resolved H$\alpha$ in emission in most, but not all epochs. We present an analysis of the H$\alpha$ morphology in \S\ref{sec:Halpha}.

\begin{deluxetable}{crr}[ht]
\tablecaption{Barycentric Radial Velocities of PSR J1306--40}
\tablehead{BJD & RV & err. \\
                   (d)  & (km s$^{-1}$) & (km s$^{-1}$)}

\startdata
2457945.5419639 & --119.5 & 6.5 \\
2457945.6301728 & --16.9 &  7.2 \\
2457956.5212405 & --123.1 & 6.8 \\
2457956.5352468 & --107.9 & 6.4 \\
2457956.5546942 & --95.8 &  8.1 \\
2457956.5687017 & --65.6 &  8.2 \\
2457956.5948590 & --57.2 &  7.0 \\
2457956.6088705 & --38.0 &  6.0 \\
2457966.4981876 & --27.2 &  7.6 \\
2457966.5122271 & --18.4 &  20.1 \\
2457966.5399208 & 38.7 &  7.2 \\
2457966.5608956 & 49.6 &   8.7 \\
2457966.5914866 & 115.1 &  10.2 \\
2457967.5998999 & --1.5 &  12.8 \\
2457996.4786139 & 231.8 &  8.4 \\
2457996.4931540 & 238.7 &  8.0 \\
2457996.5105239 & 228.8 &  9.4 \\
2458001.4795192 & --144.0 & 11.4 \\
2458001.4935388 & --124.6 & 10.2 \\
2458139.7564130 & --136.8 & 7.7 \\
2458139.7704090 & --117.1 & 6.8 \\
2458140.7400266 & --194.5 & 9.8 \\
2458140.7540222 & --166.4 & 8.3 \\
2458140.7770050 & --150.8 & 7.4 \\
2458140.7910007 & --158.5 & 8.2 \\
2458161.6721983 & --121.2 & 7.9 \\
2458161.6872977 & --107.7 & 7.3 \\
2458202.5822163 & 177.8 &  8.5 \\
2458202.5962046 & 197.7 &  9.6 \\
2458202.8350917 & 190.9 &  7.4 \\
2458202.8490793 & 176.1 &  7.6 \\
2458223.5928781 & 231.9 &  7.7 \\
2458223.8317433 & 41.2 &  8.4 \\
2458223.8456941 & 32.1 &  14.9 \\
2458243.6779211 & --87.6 &  8.6 \\
2458243.6919415 & --112.3 & 10.5 \\
2458243.7115353 & --125.5 & 9.5 \\
2458243.7263475 & --141.9 & 10.2 \\
2458243.7459698 & --130.3 & 10.9 \\
2458247.5062024 & 131.9 &  11.5 \\
2458247.5435496 & 160.5 &  14.2 \\
2458278.5472619 & 197.1 &  15.9 \\
2458278.5613912 & 183.7 &  15.7 
\enddata
\label{tab:RVdata}
\end{deluxetable}

\subsubsection{Determining the Mass Ratio}
By assuming the secondary is tidally synchronized with the  pulsar, we can estimate its projected rotational velocity ($v$ sin $i$) by comparing the rotationally broadened spectra with non-rotating template stars of similar spectral type. The assumption of synchronization is reasonable since this timescale is $\lesssim 1$ Myr for a system with the period of PSR J1306--40 and a typical redback mass ratio \citep{Zahn77}. Combined with our measurement of the orbital semi-amplitude, we use the $v$ sin $i$ value to constrain the mass ratio of the binary.

Following similar procedures as described in \citet{Strader14} and \citet{Swihart17}, we obtained spectra of bright late-G to mid-K giant stars to use as templates and convolved these with a set of rotational convolution kernels reflecting a range of $v$ sin $i$ values (including limb darkening). After cross-correlating the broadened templates with the original, unbroadened spectra, we fit a relation between the full-width at half-maximum (FWHM) values and the input value of $v$ sin $i$. We then cross-correlated the spectra of the companion to PSR J1306--40 with that of the unconvolved standard stars and used the FWHM values to estimate the projected rotational velocity.

Our final estimate of $v$ sin $i$ derived in this manner is 75.5 $\pm$ 3.8 km s$^{\rm -1}$, where the uncertainty represents the standard deviation of the measurements among all templates and does not account for systematic uncertainties. Along with our measured value of the semi-amplitude, we use this equation, which uses the Roche lobe approximation of \citet{Eggleton83}:

\begin{equation}
v \, \textrm{sin} \, i = K_2\;\frac{0.49\;q^{2/3}\;(1+q) }{0.6\;q^{2/3} + \textrm{ln}(1 + q^{1/3})}
\end{equation}

where $q = M_2/M_{NS}$ is the mass ratio. This equation  is valid assuming that the secondary fills its Roche lobe and is synchronized. We find that $q = 0.290\pm0.031$. This measurement can be directly tested once a timing solution for the pulsar is available.

%the standard formula $v$ sin $i$ = $0.462\,K_2 \, q^{1/3} \, (1+q)^{2/3}$ \citep{Casares01} to estimate the mass ratio: $q = M_{2}/M_{1} = 0.285 \pm 0.030$.

\subsubsection{The Minimum Neutron Star Mass}

The binary mass function $f(M)  = PK_2^3/2\pi G = M_{NS} \, \textrm{sin}^3 (i)/(1+q)^2$. We have determined all of these quantities from the spectroscopy alone, except for the inclination. Propagating the uncertainties appropriately, we can determine the \emph{minimum} neutron star mass  $M_{NS}$ \, sin$^3 (i)$ = $1.75\pm0.09\,M_{\odot}$. This quantity is independent of the light curve modeling we carry out in \S\ref{sec:LCfitting} and is a robust lower limit. Hence we see that, consistent with many other redbacks \citep{Strader18}, the mass of PSR J1306--40 is well in excess of the canonical $1.4\,M_{\odot}$.

\subsection{Optical Photometry}
We obtained optical $BVI$ photometry of the companion to PSR J1306--40 using ANDICAM on the SMARTS 1.3-m telescope at CTIO over 101 nights between 2017 Jul 17 and 2018 Apr 29 (UT). On each night, we took single 340 and 250 sec exposures in $V$ and $I$ bands, respectively, and two 250 sec exposures in $B$ that were merged during the reduction process. Data were reduced following the procedures in \citet{Walter12}.

We performed differential aperture photometry to obtain instrumental magnitudes of the target source using fifteen nearby comparison stars as a reference. Absolute calibration was done with respect to the \citet{Landolt92} standard fields TPheD (2017 Jul 17 -- Sep 03) and RU149 (2018 Jan 24 -- Apr 29). After excluding measurements with large errors, our final dataset includes 92, 93, and 95 measurements in $B$, $V$, and $I$ bands, respectively. The mean observed magnitudes (not corrected for extinction) are $B = 19.21$, $V = 18.35$, and $I = 17.20$, with median errors of $\sim$~0.03 mag in $V$ and $I$ and $\sim$~0.06 in $B$. The SMARTS photometry is listed in Table~\ref{tab:SMARTSdata}.

\begin{deluxetable}{cccc}[!h]
\tablewidth{170pt}
\tablecaption{SMARTS Photometry of PSR J1306--40}
\tablehead{BJD & Band & mag & err \\
                   (d) & & & }
\startdata
2457962.46060 & B & 19.104 & 0.104 \\
2457962.46769 & V & 18.212 & 0.029 \\
2457962.47235 & I & 17.007 & 0.062 \\
2457967.46273 & B & 19.357 & 0.099 \\
2457967.46981 & V & 18.631 & 0.044 \\
2457967.47448 & I & 17.461 & 0.095 \\
2457968.46610 & B & 19.334 & 0.073 \\
2457968.47319 & V & 18.426 & 0.037 \\
2457968.47785 & I & 17.289 & 0.033 \\
2457969.47164 & B & 19.140 & 0.062 \\
... & ... & ... & ...
\enddata
\tablecomments{The full SMARTS dataset is available in machine-readable format. We show a portion of the table here as a preview of its form and content. Magnitudes have not been corrected for extinction.}
\label{tab:SMARTSdata}
\end{deluxetable}

\begin{figure*}[ht]
\includegraphics[width=1\textwidth,trim={0 0 0 0},clip]{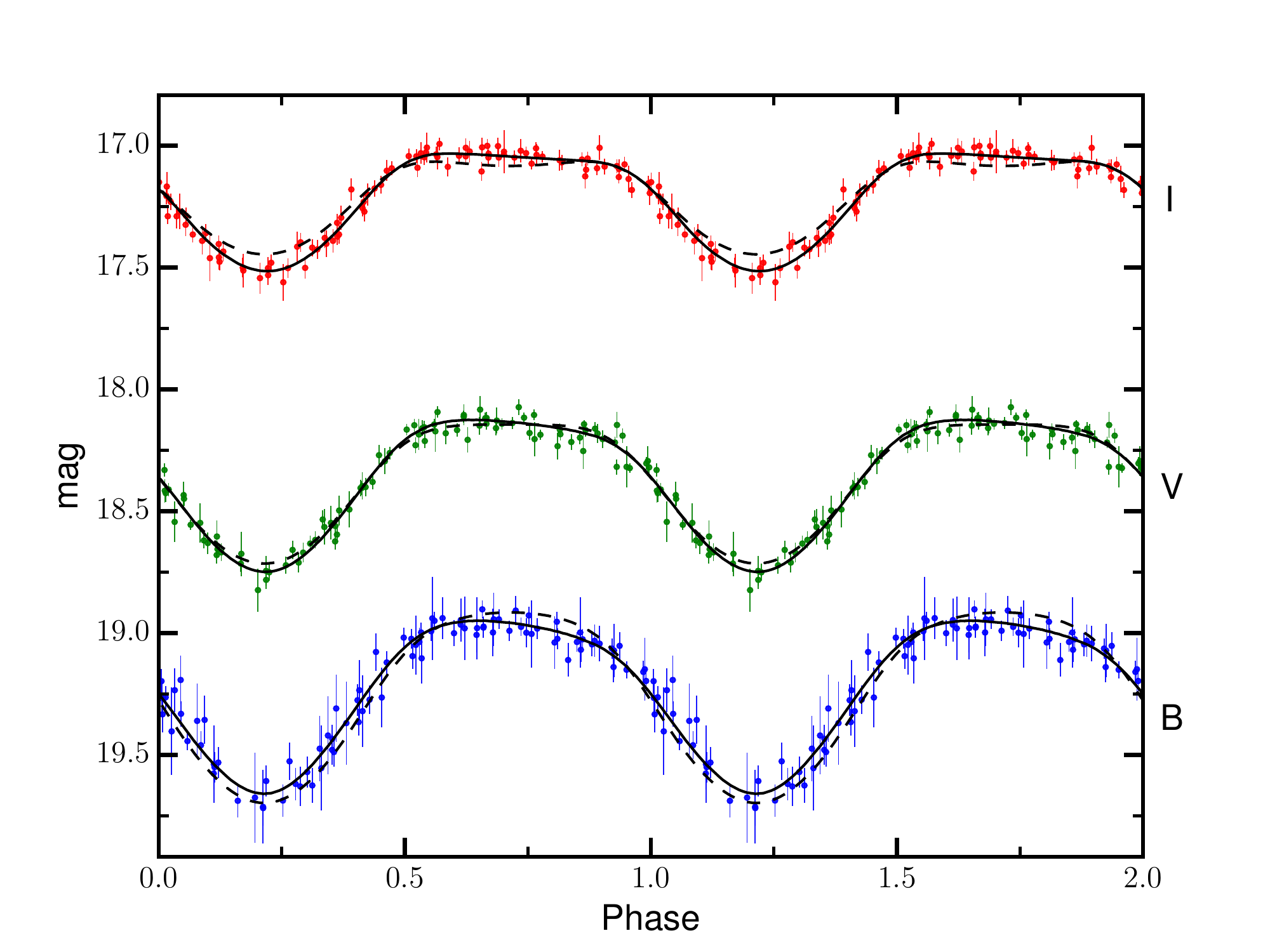}
\caption{SMARTS optical photometry of PSR J1306--40. Black lines show the best fit ``No Spot'' (dashed) and ``Hot Spot'' (solid) ELC models from Table~\ref{table:lcmodels}. Two full orbital phase cycles are shown for clarity.}
\label{fig:lcmodel}
\end{figure*}

To determine the orbital phase of these observations, we folded the data on the orbital period and ephemeris derived from our spectroscopic observations, where $\phi$ = 0 corresponds to the ascending node of the pulsar (in this convention, the secondary is between the pulsar and Earth at $\phi$=0.25). We show the folded light curves in Figure~\ref{fig:lcmodel}.

The most obvious features of the light curves are the broad maxima centered near $\phi \sim 0.75$ and narrower minima at $\phi \sim 0.25$. A maximum at this phase suggests the tidally locked ``day'' side is being heated substantially, while the minimum corresponds to the ``night'' side of the secondary as we view it at inferior conjunction. As pointed out by \citet{Linares18}, this heating plays a dominant role in shaping the optical light curve over the tidal deformation of the secondary (i.e., ellipsoidal modulations). If heating were not dominating the light curve, ellipsoidal modulations would result in two roughly equal maxima at $\phi$ = 0 and $\phi$ = 0.5, and two unequal minima at $\phi$ = 0.25 and $\phi$ = 0.75. The minimum at $\phi$ = 0.75 would be dimmer due to the effects of gravity and limb darkening when viewing along the axis connecting the primary and secondary. Overall, the shape and amplitude of the light curves appear broadly consistent with the unfiltered Catalina Sky Survey \citep{Drake09} data presented by \citet{Linares18}, suggesting no significant state change has occurred in this system since late-2005. 

Another notable feature of the light curves is that the maxima are not symmetric about the expected $\phi$ = 0.75. Instead, the light curves slope downward between 0.5 $< \phi <$ 1.0, suggesting that the source of heating is asymmetric with respect to the rotational axis of the secondary. Similar asymmetric heating has been observed in a number of other redbacks and black widows, and may be common in these systems due to the effects of an asymmetric intrabinary shock, heating mediated by the magnetic field of the secondary, or other magnetic activity such as star spots \citep[e.g.,][]{Stappers01, Breton13, Schroeder14, Romani15, Romani15b, Deneva16, Romani16, vanStaden16, Sanchez17, Cho18}. We show below that asymmetric heating is a promising (though not unique) mechanism to explain the observed light curves of PSR J1306--40.

\section{Light Curve Fitting}
\label{sec:LCfitting}
We modeled the $BVI$ light curves of PSR J1306--40 using the Eclipsing Light Curve \citep[ELC;][]{Orosz00} code. Given the recent pulsar detection, we assume there is no accretion disk; the light curves are dominated by a tidally distorted secondary that is being heated on its tidally locked day side. We also assume a circular orbit and fix the orbital period $P$, semi-amplitude $K_2$, and mass ratio $q$ to the values derived from our spectroscopy (\S\ref{sec:spectroscopy}). These values immediately set the scale for the system, giving an  orbital separation of $\sim$ 6 R$_{\odot}$.

In our most basic model, we fit for the orbital inclination $i$, the Roche lobe filling factor $f_2$, the intensity-weighted mean surface temperature of the secondary $T_{\textrm{eff}}$, the central isotropic irradiating luminosity from the pulsar, and a phase shift $\Delta \phi$. We characterize the irradiating luminosity indirectly through the most directly observed quantity: the maximum day side temperature of the heated secondary $T_{\textrm{day}}$. In Table~\ref{table:lcmodels}, we summarize this model (``No Spot'') with the medians of the posterior distributions and corresponding 1$\sigma$ uncertainties for each parameter. We also plot this model along with the data in Figure~\ref{fig:lcmodel} (dashed line). The best-fit inclination from these models is not all that well constrained ($i=70.4^{+16.7}_{-7.0}$), but implies a massive neutron star ($M_1 = 2.08^{+0.35}_{-0.34}$ \Msun) that would be among the heaviest known. For the best-fit parameters, the formal $\chi^2$/d.o.f. = 370/275, suggesting the fit can be improved. In particular, this model overestimates the $I$-band flux at $\phi \sim$ 0.25 and underestimates the $B$-band flux at this same phase, indicating the temperature profile of the night side is not well-matched. This model also does a poor job of reproducing the asymmetric, downward sloping curves between 0.5 $< \phi <$ 1.0, as the only source of heating is symmetric around conjunction.

\begin{deluxetable*}{l|cc}[!t]
\tablewidth{290pt}
\tablecaption{Summary of ELC fits for PSR J1306--40}
\startdata
\hline \hline \Tstrut
\large{Parameters} & \large{No Spot} & \large{Hot Spot} \\[3pt]
\hline
incl ($^\circ$)\Tstrut & 70.4$^{+16.7}_{-7.0}$ & 83.5$^{+3.3}_{-4.8}$\\[4pt]
filling factor ($f_2$) & 0.88$^{+0.06}_{-0.02}$ & 0.94$^{+0.04}_{-0.05}$\\[4pt]
$T_{\textrm{eff}}$ & 4728$^{+77}_{-67}$ & 4913$^{+198}_{-191}$\\[4pt]
$T_{\textrm{day}}$ & 5542$^{+275}_{-165}$ & 6187$^{+397}_{-451}$\\[4pt]
$\Delta\phi$ & -0.0354$^{+0.0016}_{-0.0019}$ & -0.0228 $\pm$ 0.0039\\[4pt]
$\lambda_{\textrm{spot}}$($^\circ$)\tablenotemark{a} & \nodata & 2$_{-2}^{+5}$\\[4pt]
$\theta_{\textrm{spot}}$($^\circ$)\tablenotemark{b} & \nodata & 27$_{-19}^{+14}$\\[4pt]
$T_{\textrm{spot}}$ (K) & \nodata & 11063$_{-414}^{+306}$\\[4pt]
$r_{\textrm{spot}}$($^\circ$) & \nodata & 52$_{-4}^{+3}$\\[4pt]
$\chi^{2}$(d.o.f.) & 370.3(275) & 220.7(271)\\[4pt]
\hline
$M_1$ (\Msun)\Tstrut & 2.08$^{+0.35}_{-0.34}$ & 1.77$^{+0.07}_{-0.03}$\\[4pt]
$M_2$ (\Msun) & 0.59 $\pm$ 0.10 & 0.51$^{+0.02}_{-0.01}$\\[4pt]
$\textrm{log}~g$ (cgs) & 3.745$^{+0.025}_{-0.010}$ & 3.725 $\pm$ 0.005
\tablenotetext{a}{Spot latitude. $\lambda_{\textrm{spot}}=0^{\circ}$ and $\lambda_{\textrm{spot}}=180^{\circ}$ represent the North and South pole, respectively}
\tablenotetext{b}{Spot longitude. $\theta_{\textrm{spot}}=0^{\circ}$ and $\theta_{\textrm{spot}}=180^{\circ}$ denote the inner Lagrange point and night side of the star, respectively}
\label{table:lcmodels}
\end{deluxetable*}

A promising solution for matching the shape of the downward sloping light curves around companion superior conjunction is by invoking an indirect, anisotropic heating model. The interaction between the pulsar and companion winds can form an intrabinary shock, generating asymmetric heating on the secondary as the spin-down power of the pulsar is reprocessed and indirectly illuminates the day side of the companion \citep{Romani16, Wadiasingh17}. In fact, a skewed intrabinary shock that slightly trails the companion during its orbit has been used (and is perhaps required) to explain the optical light curves in a number of black widow and redback MSP systems. \citep[e.g.][]{Romani15, Romani15b, Deneva16, Cho18}. Such a shock could also be a natural location for producing H$\alpha$ emission (see \S\ref{sec:Halpha}).

In addition, strong magnetic fields on the companion's surface may play an important role in the location of the anisotropic heating of the secondary. The rapid rotation and large temperature asymmetry of the tidally-locked companion can produce a strong dynamo, enhancing the surface magnetic fields \citep{Morin12}. These magnetic field lines can duct the particles accelerated in the shock, channeling them directly to the magnetic poles and creating intense localized heating (i.e., one or more hot spots) \citep{Tang14, Sanchez17}.

ELC does not allow us to directly model an offset intrabinary shock or a channeled pulsar wind. But we do have the ability to model its effect on the companion: we can add a hot spot to our underlying model of a heated, tidally distorted secondary. The hot spot is modeled as a circle with a temperature structure that falls off linearly towards the edges. We fit for the central temperature ($T_{\textrm{spot}}$), size ($r_{\textrm{spot}}$), and location ($\lambda_{\textrm{spot}}$, $\theta_{\textrm{spot}}$) of the hot spot (see Table~\ref{table:lcmodels}).

Our main result is that adding a single hot spot to the heated, distorted secondary provides an excellent fit to the light curve ($\chi^2$/d.o.f. = 221/271). The reduced $\chi^2$ in this model is $<1$ likely due to the overestimation of a subset of the photometric uncertainties. In the best fitting hot spot model, the system is highly inclined ($i\sim83^{\circ}$), with a large hot spot near the north pole of the secondary. We show this best-fit model in Figure~\ref{fig:lcmodel} (solid line) and summarize the posteriors in Table~\ref{table:lcmodels}.

Using our best hot spot model, the inferred mass of the primary is $M_1 = 1.77^{+0.07}_{-0.03}$, fairly massive for a typical neutron star, but fully consistent with the neutron star masses found in many redback MSPs \citep[e.g.,][]{Kaplan13, Romani15b, Bellm16, Strader16, Sanchez17, Shahbaz17, Linares18b, Swihart18}. The secondary has a mass of $M_2 = 0.51^{+0.02}_{-0.01}$, placing it in the high-mass tail of the redback companion mass distribution \citep{Strader18}. The inferred gravity of the companion is $\textrm{log g} = 3.725 \pm 0.005$ (cgs), suggesting it is slightly evolved off the main sequence, which may be relevant for interpreting the high energy emission from this system in the context of magnetically driven winds/outflows.

For the hot spot model, we require a small phase shift to obtain adequate fits to the light curves. This is primarily due to the light curve minimum occurring $\sim$30 minutes earlier than what we expect from our spectroscopic ephemeris. This shift is larger than what we can account for from our formal uncertainties on the orbital period and $T_0$, and the maxima appear roughly centered on the expected $\phi=0.75$, so this feature of the light curves is likely to be real. One interpretation of this feature is that in addition to the hot spot, the intrabinary shock could be slightly trailing the companion, leading to off-center heating.

For the hot spot itself, echoing our previous discussion, its best-fit location is centered near the companion's north pole, consistent with magnetic ducting of intrabinary shock particles. The spot covers roughly $\sim 19\%$ of the surface area of the star, but contributes about a third of the observed $V$-band flux owing to its high temperature. This value is broadly consistent with the contribution from a similar hot spot modeled for the optical light curves of the redback binary 3FGL J0212.1+5320 \citep{Shahbaz17}.

\subsection{Distance}
\label{sec:distance}
Using the results from our light curve models, we estimate the distance to the binary by comparing its apparent magnitude to its intrinsic luminosity following similar procedures outlined in \citet{Strader15} and \citet{Swihart18}. We estimate the bolometric luminosity by assuming $T_{\textrm{eff}}$ and the inferred radius of the secondary ($R \sim 1.6\,R_{\odot}$) from our best-fit model (Table~\ref{table:lcmodels}, column 3). We fit 10 Gyr solar metallicity isochrones \citep{Marigo08} to estimate bolometric corrections and used these to obtain the predicted absolute magnitude in each band ($BVI$). Finally, we compared these values to the mean apparent magnitudes after applying extinction corrections using the \citet{Schlafly11} reddening maps to get a distance to the binary: 4.7 $\pm$ 0.5 kpc. For the remainder of this paper, we adopt this value for the distance. We note that this method has proven successful in the past at estimating reliable distances to similar systems \citep[e.g.,][]{Swihart18}. However, due to systematic effects such as the unknown metallicity and precise evolutionary state of the star, our estimate is likely uncertain by at least 20\%. Future \emph{Gaia} releases will help us address these systematics.

PSR J1306--40 is listed in \emph{Gaia} DR2\footnote{https://www.cosmos.esa.int/web/gaia/dr2} with parallax 0.106 $\pm$ 0.278 mas. Although the current parallax measurement is not significant, we estimate the geometric distance to the binary by using a weak distance prior based on an exponentially decreasing space density Galaxy model with a scale length of 0.94 kpc \citep{BailerJones18}. We also calculated the distance using a larger scale length (1.35 kpc), suggested by \citet{Astraatmadja16}. The resulting distances are 2.83$^{+1.65}_{-0.99}$ kpc and 3.47$^{+2.37}_{-1.36}$ kpc, respectively, somewhat smaller than our optically-inferred distance but within the uncertainties.

Although the above estimates are uncertain, they are all wholly inconsistent with the dispersion measure based distance estimates made by using the measured dispersion measure and the \citet[][CL02]{Cordes02} or \citet[][Y17]{Yao17} electron density models ($d_{\textrm{CL02}}$$\sim$1.2 kpc and $d_{\textrm{Y17}}$$\sim$1.4 kpc, respectively).

The discrepancies between the light curve/parallax distances and the dispersion measure-based estimates are consistent with the results of \citet{Jennings18}, who use \emph{Gaia} DR2 parallaxes to measure the distances to a number of black widows and redbacks, and find that some dispersion measure distances, particularly those at high Galactic latitude, are underestimated. This also agrees with previous results for the distances of normal MSPs outside the Galactic Plane \citep{Gaensler08, Roberts11b}.

\section{H$\alpha$ Spectroscopy}
\label{sec:Halpha}
A number of MSP binaries have shown strong H$\alpha$ emission lines. In some of these systems, the existence of double-peaked Balmer emission indicates the presence of an accretion disk \citep[e.g.,][]{Martino14, Bogdanov15}, although an intrabinary shock and/or material streaming off the companion have also been used to explain the complex morphology \citep{Sabbi03, Swihart18}. Here we present the phase-resolved H$\alpha$ profiles and suggest an intrabinary shock near the companion's surface can explain the origin of the emission.

\begin{figure}[ht]
\includegraphics[width=1\linewidth]{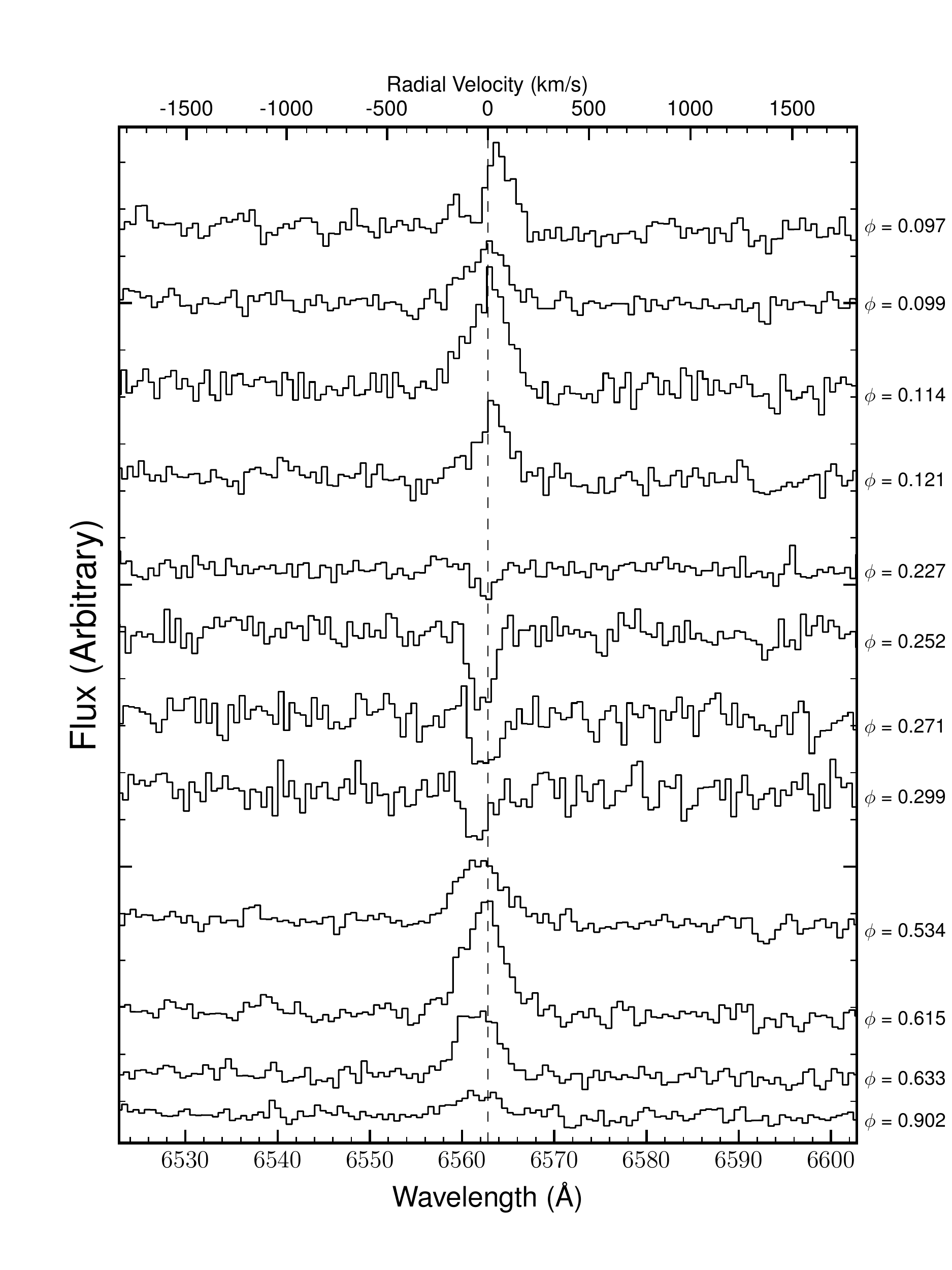}
\caption{Phase-resolved optical spectra of PSR J1306--40 around the H$\alpha$ region. Spectra have been corrected to the rest frame of the secondary (dashed line) and arbitrarily shifted in flux to display changes in the profile shape. The switch from emission to absorption near companion inferior conjunction ($\phi = 0.25$) is apparent, likely due to an eclipse of the emission region.}
\label{fig:halpha_fig}
\end{figure}

Throughout most of our spectroscopic monitoring of PSR J1306--40, a moderate H$\alpha$ emission line was persistent, occasionally showing complex, double-peaked morphology. At other times, the H$\alpha$ appears only in absorption. We show the phase-resolved H$\alpha$ profiles in Figure~\ref{fig:halpha_fig}. Each spectrum has been corrected to the rest frame of the secondary (dotted line) and arbitrarily shifted in flux to display changes in the profile shape.

When in emission, the H$\alpha$ profile always peaks directly at, or very near, the velocity of the secondary. This suggests the emission is either coming directly from the star or from a region relatively close to the companion's surface. Since the emission tracks the orbital motion, the line could arise from the stellar chromosphere as is the case in RS CVn systems \citep[e.g.,][]{Drake06}. However, the emission we observe is broad, with a significant amount of the emission present at velocities $\gtrsim$200 km s$^{\textrm{-1}}$ from the radial velocity of the companion. Furthermore, near companion inferior conjunction ($\phi$=0.25), the emission line disappears and instead only appears in absorption. Since H$\alpha$ absorption is typical in low-mass stars like the companion to PSR J1306--40 \citep{Cram85, Pickles98}, it is likely that the emission region is being eclipsed at these phases and we are only seeing H$\alpha$ absorption intrinsic to the star.

Such a scenario can be explained by invoking an H$\alpha$ emitting intrabinary shock near the companion's surface. Given our evidence for a rapidly-rotating Roche-lobe filling companion, it is likely the magnetic field on the surface is enhanced, which can lead to strong winds from the likely subgiant secondary. These winds can then be heated directly by the pulsar's radiation pressure or through an interaction with the pulsar wind as it forms an intrabinary shock (see \S\ref{sec:LCfitting}). Given the orbitally-modulated X-ray emission found by \citet{Linares18}, such a shock almost certainly exists in this system and can provide a natural explanation for the observed X-ray and H$\alpha$ phenomenology.

\section{Discussion}
\label{sec:discussion}
We have presented new optical photometry and spectroscopy of the source PSR J1306--40, a redback-like MSP binary with one of the longest known orbital periods among MSPs with non-degenerate companions. Our modeling suggests the companion is somewhat evolved and massive compared to typical redbacks, which may be relevant for interpreting the high-energy emission in the context of an intrabinary shock. 

As pointed out by \citet{Linares18}, PSR J1306--40 lies in the error region of the \emph{Fermi}-LAT $\gamma$-ray source 3FGL J1306.8--4031 \citep{Acero15}, about 3.4\arcmin~from its center. In the preliminary LAT 8-year point source catalog (FL8Y\footnote{https://fermi.gsfc.nasa.gov/ssc/data/access/lat/fl8y/}), which includes four additional years of survey data, the updated error region corresponding to the source FL8Y J1306.8--4035 is now $\sim$60\% smaller in area than in 3FGL and has been shifted $\sim$3.2\arcmin~South, such that PSR J1306--40 lies only $\sim$0.86\arcmin~away from its center. Hence there is compelling evidence that PSR J1306--40 is indeed associated with a \emph{Fermi}-LAT $\gamma$-ray source.

Although the $\gamma$-ray spectra of many MSPs are typically highly curved, the relatively flat 3FGL spectrum shows no significant evidence for curvature with no cutoff up to $\gtrsim$10 GeV, reminiscent of huntsman MSP 1FGL J1417.7--4407 \citep{Strader15,Camilo16}, which has a red giant secondary in a 5.4 day orbit and also shows a flat $\gamma$-ray spectrum. Similar to PSR J1306--40, 1FGL J1417.7--4407 likely has a strong intrabinary shock and a significant wind from the companion that may be influencing $\gamma$-ray production \citep{Swihart18}. \citet{Linares18} attributes the unusual $\gamma$-ray spectrum of 3FGL J1306.8--4031 to contamination from two nearby, active galaxies. We do not rule out this possibility, but given the improved source localization in FL8Y, contamination from other sources may not be important. The upcoming official 4FGL catalog will allow a reassessment of the spectrum of the $\gamma$-ray source.

The X-ray light curve shows one maximum and one minimum per orbital period \citep{Linares18}, which can be attributed to emission from an intrabinary shock that becomes at least partially eclipsed during inferior conjunction of the companion \citep[e.g.,][]{Bogdanov11,Rivera18}. Although the minimum X-ray count rate presented by \citet{Linares18} approaches zero, there is no clear evidence for a total eclipse of the X-ray emitting region. At our best fit inclination ($i=83.5^{\circ}$), an X-ray emitting point source would be completely eclipsed between $\phi \sim 0.21-0.29$, suggesting the shock is somewhat extended. Additional data will be needed to probe the precise geometry of the shock.

\citet{Linares18} estimated the epoch of inferior conjunction of the companion from fitting the time of minimum of the X-ray light curve ($T_{0,\textrm{X}}$). Comparing this value with our spectroscopic ephemeris, we find the X-ray light curve lags behind the companion orbit by about one hour; the X-ray flux reaches a minimum at $\phi \sim 0.29$, with a similar offset needed to match the X-ray maxima. However, $T_{0,X}$ and our spectroscopic $T_0$ are separated in time by 1471 d, approximately 1340 orbital periods. Assuming our uncertainty on the binary period (0.000161 d), building a phase-coherent solution for the radial velocity data backward to the epoch of the X-ray observation results in an absolute uncertainty of $\sim$12.9 hours, nearly half an orbital period. Therefore, the lag we find is not reliable with the current data. In the future, a precise pulsar timing solution would enable us to check the relative phase alignments between the X-ray and optical light curves.

PSR J1306--40 has a long orbital period and a relatively massive secondary compared to the population of known redbacks in the Galactic field \citep{Strader18}. In the binary evolution models of \citet{Podsialowski02}, PSR J1306--40 lies above the bifurcation period that distinguishes systems whose orbits will shrink to become black widows/redbacks and those that will grow to become MSP--white dwarf binaries. The light curves of PSR J1306--40 also show strong signs of irradiation, which can further contribute to an increase in the orbital period if the companion undergoes even low levels of evaporation \citep{Chen13}. Our light curve models also suggest that the companion is filling a substantial fraction of its Roche lobe and that it has a somewhat evolved, subgiant-like radius, luminosity, and gravity ($\textrm{log g} = 3.725 \pm 0.005$). Given that PSR J1306--40 has been spun up to obtain its rapid spin period, this would put PSR J1306--40 on a standard Case B evolutionary track of low-mass X-ray binaries that started the mass transfer process after leaving the main sequence, placing it in the late stages of the MSP recycling process that will terminate with a wide orbit MSP--white dwarf binary \citep{Tauris99}. 

The relatively long period and subgiant-like secondary of PSR 1306--40 resemble the huntsman systems 1FGL J1417.7--4407 and 2FGL J0846.0+2820 \citep{Strader15,Camilo16,Swihart17,Swihart18}, which are likely progenitors to the typical MSP--He white dwarf binaries observed in the Galactic field. Another similarity between PSR 1306--40 and at least 1FGL J1417.7--4407 is the X-ray luminosity: \citet{Linares18} inferred a 0.5--10 keV X-ray luminosity of $8.8 \times 10^{31}$ erg s$^{\textrm{-1}}$ for PSR 1306--40 when using the dispersion measure-based distance of 1.2 kpc. But at our optical light curve-inferred distance estimate of 4.7 kpc, the X-ray luminosity of PSR 1306--40 is $\sim$10$^{33}$ erg s$^{\textrm{-1}}$,  brighter than nearly all known redbacks in the pulsar state---except 1FGL J1417.7--4407.

Detecting radio pulsations in these huntsman systems has proven difficult even compared to typical redbacks \citep{Camilo16, Keane17}. This difficulty may be related to the strong winds from the evolved companions, which when ionized could eclipse the radio emission more readily than for redbacks with main sequence-like companions.

This high inferred eclipse fraction, and the rarity of these systems due to the shorter length of the red giant phase of evolution compared to the main sequence phase, suggests discovering other huntsman systems may be even more reliant on multiwavelength follow-up of unassociated \emph{Fermi}-LAT error regions than for typical redbacks. Fortunately, the companions in these systems are intrinsically brighter than redbacks, and so more readily observable in ground-based optical variability studies.

A precise pulsar timing solution would be a significant step for fully understanding the huntsman PSR J1306--40 and its connection to known redbacks, placing tighter constraints on the orbital dynamics and permitting a search for spin and/or orbitally modulated $\gamma$-ray pulsations in the \emph{Fermi} data. Additional deep X-ray observations would also be useful to compare with the most recent optical light curves and with detailed intrabinary shock models. The subgiant nature of the nearly Roche lobe filling, highly irradiated companion, and the clear evidence for intrabinary material makes this system a good candidate for future multiwavelength monitoring.

\section*{Acknowledgements}
We thank an anonymous referee for helpful comments that improved the paper. We also thank M. Linares for useful discussions, and J. Orosz for assistance with the ELC code. We gratefully acknowledge support from NASA grant 80NSSC17K0507 and National Science Foundation grant AST-1714825. JS thanks the Packard Foundation for support.

Based on observations obtained at the Southern Astrophysical Research (SOAR) telescope, which is a joint project of the Minist\'{e}rio da Ci\^{e}ncia, Tecnologia, Inova\c{c}\~{o}es e Comunica\c{c}\~{o}es (MCTIC) do Brasil, the U.S. National Optical Astronomy Observatory (NOAO), the University of North Carolina at Chapel Hill (UNC), and Michigan State University (MSU).

\bibliography{report}

\end{document}